# Identification and super-resolution imaging of ligand-activated receptor dimers in live cells

*Pascale Winckler[1,2], Lydia Lartigue[3], Gregory Giannone[4,5], Francesca De Giorgi[6], François Ichas[6], Jean-Baptiste Sibarita[4,5], Brahim Lounis[1,2] and Laurent Cognet[1,2,*]*

[1]*LP2N, University of Bordeaux, UMR 5298, F-33405 Talence, France*
[2]*Institut d'Optique & CNRS, UMR 5298, F-33405 Talence, France*
[3]*INSERM U916 VINCO, Institut Bergonié, University of Bordeaux, F-33076 Bordeaux, France*
[4]*Interdisciplinary Institute for Neuroscience, University of Bordeaux, UMR 5297, F-33000 Bordeaux, France*
[5]*CNRS, Interdisciplinary Institute for Neuroscience, UMR 5297, F-33000 Bordeaux, France*
[6]*Fluofarma, Pessac, France*

*\*Correspondence to lcognet@u-bordeaux1.fr*

**Abstract**

Molecular interactions are key to many chemical and biological processes like protein function. In many signaling processes they occur in sub-cellular areas displaying nanoscale organizations and involving molecular assemblies. The nanometric dimensions and the dynamic nature of the interactions make their investigations complex in live cells. While super-resolution fluorescence microscopies offer live-cell molecular imaging with sub-wavelength resolutions, they lack specificity for distinguishing interacting molecule populations. Here we combine super-resolution microscopy and single-molecule Förster Resonance Energy Transfer (FRET) to identify dimers of receptors induced by ligand binding and provide super-resolved images of their membrane distribution in live cells. By developing a two-color universal-Point-Accumulation-In-the-Nanoscale-Topography (uPAINT) method, dimers of epidermal growth factor receptors (EGFR) activated by EGF are studied at ultra-high densities, revealing preferential cell-edge sub-localization. This methodology which is specifically devoted to the study of molecules in interaction, may find other applications in biological systems where understanding of molecular organization is crucial.

# INTRODUCTION

By providing optical images with spatial resolutions below the diffraction limit, super-resolution fluorescence microscopies opened the possibility to study biological structures with finer details compared to conventional light microscopies[1,2]. Most existing methods rely on the optical control of the nano-emitters fluorescent state population. For instance, localization-based methods consist in recovering the positions of numerous single molecules with high accuracy from image sequences where each frame contains only a few stochastically photoactivated emitters[3-6]. Several of these methods are suitable for imaging live bio-samples[7-9], giving unique details about spatio-temporal molecular organizations at the scale of a few tens of nanometers in cells[10,11]. Functional signaling molecules are frequently highly organized in molecular assemblies. Molecular interactions occur at the nanometer scale and are generally highly dynamic. They are often triggered by an external activating signal like a specific ligand binding. However, although of prime importance to unravel several key molecular processes, the identification in super-resolved imaging of molecular assemblies like multimers, is still lacking. Here we demonstrate that ligand activation can serve as the stochastic process necessary for identification and localization-based super-resolution microscopy of receptor dimers. To this aim, we continuously generated single-molecule FRET signals originating from dimerized receptors by using distinct fluorescent ligands and recording their receptor binding in real-time, This allowed obtaining for the first time specific images of ligand-activated molecular dimers with super-resolution and further study their dynamic behavior in the early stage of their activation. We chose epidermal growth factor receptors (EGFR) as model system. We reveal specific membrane localization of EGFR dimers upon binding of epidermal growth factor (EGF). In addition we could specifically track the movement of individual receptor dimers with unprecedented statistics on a single cell and compare their dynamic behavior to that of the full EGFR population.

EGFRs are cell-surface receptors belonging to the ErbB family of receptor tyrosine kinases and are activated following binding to their extracellular domain with members of the epidermal growth factor (EGF) family. Because of the EGFR signaling significance in different carcinoma types, EGFR has been the focus of several studies using fluorescent microscopy at both ensemble[12-15] and single-molecule levels[16-19]. Like many signaling receptors, EGFRs are widely expressed at the membrane in monomeric and dimeric states. Binding of EGF to receptors indeed induces receptor dimerization and tyrosine autophosphorylation that activates intracellular signaling cascades leading to phenotypic changes like increased proliferation and migration. Importantly, receptor endocytosis occurs in the minute time scale following EGF binding[20] such that activated EGFRs dimers are only transiently present at the plasma membrane making their study challenging at high resolution in living cells.

**RESULTS**

**Live cell super-resolution imaging of functional membrane EGFRs**

The observation of ligand-activated membrane EGFR dimers by single molecule FRET requires first that specific imaging of functional EGFRs newly activated by EGF can be obtained with high resolution at the cell membrane. To this aim, we designed a two-color super-resolution microscope based on the principle of uPAINT[6,9]. uPAINT relies on stochastically labeling in real-time of target biomolecules by fluorescent probes, and simultaneously recording their localization and dynamics on the cell membrane at the single molecule level using oblique illumination[21] (Fig. 1a). The two-colors optical setup is built around a home-made dual-view system operating with single charge-coupled device camera[22] (Methods and Supplementary Fig.1). In a first experiment, live COS7 cells starved from growth factors overnight are illuminated with a 532nm laser beam. Immediately after the beginning of recording, fluorescent EGF-Atto532 is introduced at low concentration (0.4 nM) in the imaging solution. Fluorescence images (8000 consecutive CCD frames) are recorded in the green detection channel with an integration time of 50 ms. The excitation beam angle is set to produce an inclined sheet of light above the glass slide[23] with an illumination thickness of $\sim 2\mu m$ at the center of the field of view. In those conditions, ligands newly bound to their target membrane receptor are efficiently illuminated while unbound fluorescent molecules freely diffusing in solution are mainly not illuminated. In addition, unbound molecules diffusing close to the cell membrane spend statistically at most two consecutive frames in the oblique illumination beam[9]. Such unwanted events are rejected from the analysis performed with homemade software (Methods). Continuous labeling and bound ligand photobleaching ensures sparse single molecule detection at the surface of the cells in each camera frame. Single molecule localizations are obtained with sub-diffraction precisions following image analysis (Methods). Fig. 1b shows a reconstructed image of endogenous EGFRs activated by EGF-Atto532 binding. The image is reconstructed from $1.6 \ 10^5$ EGFR localizations belonging to ~$10^4$ single molecule trajectories (Supplementary Fig.2). Functionality of fluorescent ligand was controlled by observing that EGFR internalization occurs within minutes following binding of fluorescent EGF[20] (Supplementary Fig.3).

Noteworthy, the detection of each individual EGFR starts at the time an EGF binds the receptor. Thus an EGFR has to be at the membrane to be accessed, excluding any receptor complex localized just beneath the membrane in an early endosome. Receptor activation is captured with one imaging frame resolution (50ms) and the detection of the activated receptors lasts until the fluorophore photobleaches (typically in one second, see Supplementary Fig. 4). In our imaging conditions, this bleaching time being shorter than the lifetime of the activated receptors at the membrane before endocytosis[20], the super-resolution images exclusively display EGFRs that are present at the membrane. Capturing receptors in their early states following ligand binding would not be possible with photo-activation based super-resolution methods since fluorophore photoactivation and ligand binding processes are not time-correlated.

We next designed a live cell competition assay to evaluate the specificity of EGF-Atto532 labeling. We used panitumumab, a human monoclonal antibody highly specific to EGFRs which impedes EGF binding[24] (Supplementary Fig.5). When a red fluorescent dye, Atto-647N excited with a 633 nm He-Ne laser was coupled to panitumumab (see methods), fluorescent panitumumab binding to EGFRs are detected in the red imaging channel producing uPAINT super-resolved images (Fig. 1c) akin to EGF labeling (Fig. 1b). Membrane EGFR localizations differences might in principle be found between antibody-labeled EGFRs and EGF-labeled EGFRs since panitumumab prevents activation of the receptors while EGF activates the receptors. However, revealing such differences would require a specific study which is out of the scope of this work. In addition, if a uPAINT acquisition is started with EGF-Atto532 as in Fig. 1b and panitumumab is added in excess after a few seconds of recording (supplementary Fig. 6), then a dramatic drop of the number of fluorescent EGFR detected is observed (Fig. 1d). All together, these experiments indicate that fluorescent EGF labeling is highly specific and allow imaging functional and newly activated EGFRs with high resolution.

**Live cell super-resolution imaging of dimers of ligand-activated EGFRs**

The super-resolved image presented in Fig. 1b displays the entire EGFR population localizations found at the membrane of live cells immediately after EGF activation, without distinction about the monomeric or multimeric state of the receptors. In particular, although present, EGFR dimers cannot be distinguished from isolated receptor in such images. In order to obtain images of EGFR dimers with super-resolution, we present in the following FRET experiments performed withEGF-Atto532 and EGF-Cy5 introduced in the imaging medium at equal concentration (~0.2 nM). We chose Atto532 and Cy5 as a FRET pair, for their relatively large Forster radius[25], estimated to ~65 Å. In order to identify FRET events (Fig. 2a), a single excitation laser (532nm) was used and the images of the green and red camera channels were simultaneously recorded. In these illumination conditions, EGF-Atto532 is excited efficiently and detected solely in the green channel while EGF-Cy5 does not produce detectable signals when used alone (Supplementary Fig. 7). Gathering single molecule localizations recorded in the donor channel a super-resolved image of EGF activated EGFRs is obtained (Fig.2b) (from 42,541 localizations corresponding to 7,078 single molecule trajectories), giving similar information than in Fig. 1b where EGFR monomers and multimers could be distinguished. Importantly, fluorescent spots are also recurrently observed on the cell surface in the acceptor channel. They originate from single molecule FRET occurring between EGFR dimers activated by an EGF-Atto532 and an EGF-Cy5 (Fig.2a). This is further evidenced by the observation of anti-correlated signal detections in corresponding positions of the donor and acceptor channels (Fig.2c and Supplementary Fig. 8). Collecting the single molecule localizations obtained by image analysis in the acceptor channel, super-resolved images of EGF activated dimer EGFRs are reconstructed as shown on Fig.2d (from 18,481 localizations corresponding to 3,350 trajectories) and on Supplementary Fig. 9. Interestingly, the content of the acceptor channel is exclusively constituted by the subpopulation of EGFR dimers activated by two EGF molecules (an EGF-Atto532 and an EGF-Cy5). In the donor channel, this subpopulation was also present (in the form of dimers labeled by two EGF-Atto532) but was undistinguishable from other activated EGFRs populations (including EGFR monomers). It is thus possible to compare the images reconstructed from the two channels in order to extract the specific localization of newly activated EGFR dimers from the entire population. It is clearly seen that they are preferentially found at the cell edge (Fig. 2 b-c). This specific localization of EGFR dimers also reported in other cell lines with low density single molecule imaging techniques[18] reminds the notion that EGFR signaling can be regionalized and causes the recruitment of several intracellular signaling proteins involved in cell motility, especially at the leading edge of the cell[26]. This is in line with recent observations describing that EGFR is particularly activable in membrane regions with a high curvature[27]. This regionalization might also be related to preferential cell edge EGF activation of Ral proteins (from the RAS-family GTPase) observed in COS 7 cells[28]. Indeed, Ral binding proteins (such as RalPB1) where proposed to regulate endocytosis of EGFRs via the partner protein of RalPB1, POB1[29].

**Membrane diffusion properties of ligand activated EGFRs dimers**

We next used single molecule tracking to study the diffusion properties of EGFR at the cell membrane. Dimers mobility could thus be compared to that of the mixed population found in the donor channel with high statistics. We analyzed the trajectories lasting more than 200ms detected in the acceptor (n= 2025) and donor (n=4530) channels. Examples of such trajectories are displayed in Fig.3a-b. We computed the mean square displacement and measured the slopes at the origin to extract the instantaneous diffusion coefficient, D, of each tracked entity[30] (Methods). Molecules with diffusion constant $<7 \times 10^{-3} \mu m^2/s$ were considered as immobile within our resolution[30]. Fig.3c presents the cumulative distribution of D values obtained on a single cell for EGFR dimers and for the entire population of activated EGFRs imaged in the donor channel. Both distributions present an heterogeneity of diffusion coefficients (ranging from highly mobile to immobile molecules) suggesting that the heterogeneity observed in the donor channel (green data point in Fig. 3c) is not primarily due to the multiplicity of EGFR multimeric compositions. Finally, the proportion of immobile dimers for the pure dimer population found by FRET (red data points in Fig. 3c) is more pronounced than for the entire population of activated EGFRs imaged in the donor channel. Activated EGFR monomers are thus likely to be more mobile than activated dimers. The frequent immobilizations undergone by the dimer population might be in part the consequence of dimer trapping in preformed endocytotic coated pits[31].

**DISCUSSION**

In this work, we have demonstrated super-resolution imaging and tracking of endogenous EGFR dimers using single molecule FRET in living cells. Based on two-color uPAINT, our method bears the key feature to rely on the stochastic real-time imaging of ligand binding and not on fluorophore photo-control of formerly labeled samples as in other super-resolution methods. This ensures that newly activated receptors are continuously captured. High-density single molecule

FRET is thus obtained allowing transient molecular interactions to be extracted and specifically studied. Our method may be applied to the study of other molecular assemblies than EGFR in live or fixed samples owing to the advance of site specific labeling of proteins with small fluorescent molecules[32,33]. For instance those developed with click chemistry tools[34] hold great promise toward this end since they have demonstrated efficient single molecule FRET capabilities[35]. This new approach giving access to the study of molecular interactions with both high resolution and high statistics should thus find applications in different systems of biology where understanding of complex nanoscale molecular organization is crucial.

**METHODS**

*Labeling of the EGF and antibody with fluorescent dyes*

Recombinant mouse EGF (R&D systems) was conjugated with Atto532-NHS-ester (Atto-Tec) or Cy5-NHS-ester (Amersham Bioscience) by using modified versions of the manufacturers' procedures. Briefly, 100 $\mu$g of EGF (1mg/mL) were incubated with 10 $\mu$L of Atto532-NHS-ester (or Cy5-NHS-ester) (5mM) in the presence of NaHCO3 (0.1 M) for 2 h at room temperature. Separation of labeled ligands from unbounded dyes was performed in size-exclusion columns (Sephadex G25; Pharmacia, New Market, NJ). As mouse-EGFs have only one reactive amino residue, the protein/dye ratio was 1:1. Panitumumab antibodies were labeled with Atto647N-NHS-ester (Atto-Tec) using the same protocol (antibodies:dye ratio of about 1:1).

*Sample preparation*

COS 7 cells are cultured in DMEM (Gibco) with 10% FBS. The day before the experiment, cells are detached with trypsin/EDTA and platted on clean coverslips. After few hours, cells are washed and cultured in serum free condition. Experiments are performed in Ringer (in mM: 150 NaCl, 5 KCl, 2 CaCl$_2$, 2 MgCl$_2$, 10 HEPES, 11 Glucose, pH 7.4) with 1mg/mL bovine serum albumin to reduce non-specific ligand adsorption. Before acquisition, the sample is incubated with a solution containing a low concentration of fluorescent beads to provide upon unspecific adsorption on the coverslip immobile reference objects used to correct long-term mechanical instabilities of the microscope. Then, the coverslip is mounted on an open chamber and 300 $\mu$l of Ringer solution is added onto the cells. At the beginning of the camera recording, 10 $\mu$L of fluorescent ligands are added (final concentration is about 0.4nM).

*Two-color uPAINT setup*

uPAINT acquisitions are performed on a custom-made dual-color microscope (Figure S1). It is based on an inverted microscope (Olympus) equipped with a 100x 1.45NA objective. Fluorophores are excited in wide-field oblique illumination by tilting a collimated laser beam in the object focal plane of the imaging lens which focuses the beams in the back focal plane of the objective. The resulting angle at the sample is set to ~5°. This allowed to image individual fluorescent ligands which have bound to the cell surface while not illuminating the molecules in the above solution[9,21]. The angle was chosen to obtain an illumination thickness of ~ 2$\mu$m in the center of the field.

Atto532 dyes were excited by a 532 nm laser solid state laser (Compass 415M, Coherent). Cy5 dyes or Atto647N dyes were excited by a 633 nm HeNe (Thorlabs). A dichroic filter (Semrock FF655-Di01) placed in the infinity detection path combined with a double bandpass emission filter (Semrock FF01-577/690) allows a spectral selection of the fluorescence signals in the donor (Atto532) and acceptor (Cy5) channels obtained simultaneously in two separated images on the EM-CCD camera (QuantEM512SC, Photometrics). A slit (~5 mm width) is placed in the imaging plane of the tube lens to avoid overlap of the donor and acceptor images on the CCD chip. Images are acquired at 20 frames/s rates. Excitation intensities were ~2kW/cm$^2$ and ligand concentrations were adjusted in order to have a constant pool of ~ 0.5 mol/$\mu$m$^2$ fluorescent molecules. We used fluorescent beads adsorbed on the glass coverslips as immobile fiduciary markers to correct for long-term mechanical instabilities of the microscope.

Control uPAINT experiments with EGF-Cy5 alone showed no single molecule detections in either detection channel using 532 nm laser excitation. In addition, when EGF-Atto532 is introduced alone, no single molecule detection can be detected in the red channel.

*Single molecule segmentation and tracking*

A typical single cell, acquired with the uPAINT microscope setup and protocol described above, leads to a set of 8,000 images that further need to be analysed in order to extract molecule localization and dynamics. Single molecule fluorescent spots were localized in each image frame and tracked over time using a combination of wavelet segmentation[36] and simulated annealing algorithms[37,38]. Under the experimental conditions described above, the resolution of the whole system was quantified to ~ 40 nm (full width at half maximum). Super-resolved images were computed by cumulating detections for all frames, using the same intensity for each localization. To analyze the trajectories we used the mean squared displacement $MSD$ computed as :

$$MSD(t = n \cdot \Delta t) = \frac{\sum_{i=1}^{N-n}(x_{i+n} - x_i)^2 + (y_{i+n} - y_i)^2}{N - n}$$

where $x_i$ and $y_i$ are the coordinates of the label position at time $i \cdot \Delta t$. We defined the measured diffusion coefficient $D$ as the slope of the affine regression line fitted to the $n = 1$ to 4 values of the $MSD(n \cdot \Delta t)$. Short-trajectories (<4 points), were filtered out. Immobile trajectories were defined as trajectories with D<0.007 $\mu m^2.s^{-1}$, corresponding to molecules which explored an area inferior to the one defined by the image spatial resolution (~0.05$\mu m$)$^2$ during the time used to fit the initial slope of the MSD.

## ACKNOWLEDGEMENTS

We thank Dr. Stéphane Pedeboscq for the generous gift of antibodies (Panitumumab). We acknowledge financial support from the Agence Nationale de la Recherche, Region Aquitaine, the French Ministry of Education and Research, the European Research Council and FranceBioImaging (Grant N° ANR-10-INSB-04-01)

## AUTHOR CONTRIBUTIONS

PW, GG, BL and LC developed the 2-color uPAINT set-up. JBS developed the analytical tools. PW and LFL performed the experiments. PW and JBS performed the analysis. FDG and FI proposed EGFR as a relevant biological model. LC conceptualized the method. BL and LC coordinated the study. PW, BL and LC wrote the manuscript. All authors discussed the results and commented on the manuscript.

## ADDITIONAL INFORMATION

The author(s) declare no competing financial interests

## FIGURES CAPTIONS

**Fig. 1. Live cell super-resolution imaging of functional membrane EGFRs newly activated by their ligand.** (a) Principle of the super-resolution method. Oblique illumination (light green) does not excite EGF ligands in solution. (b) uPAINT image of EGFR labeled by EGF-Atto532 acquired on live cells. (c) Same experiment performed using Panitumumab-Atto647N. (d) Competition assay showing specificity of EGF-Atto532 labeling: number of fluorescent EGF detected per frame (50 ms) on the cell membrane during a uPAINT acquisition using EGF-Atto532. After ~ 8s and 38s (red arrows), unlabeled Panitumumab was added in 100-fold excess (40 nM) compared to EGF.

**Fig. 2. Live cell super-resolution imaging of membrane EGFR dimers based on single-molecule FRET.** Dual color uPAINT imaging of EGFR was performed using a 1:1 mix of EGF-Atto532 and EGF-Cy5 under 532 nm laser excitation. (a) Schematics of single molecule FRET between two fluorescent ligands bound on a EGFR dimer. (b) Donor channel: super-resolved image of EGFR labeled by EGF-Atto532 as in Fig. 1b. (c) Acceptor channel: super-resolved image of EGF activated dimer EGFRs obtained by single molecule FRET. (d) Signature of single molecule FRET: anti-correlated fluorescence signals detected by single molecule fitting in the donor (green line) and acceptor (red line) channels, in corresponding positions. Insets in (b) and (c) represents zooms of highlighted regions showing preferential cell edge localization of the dimers.

**Fig. 3. Membrane dynamics of EGFR dimers based on single molecule tracking of the FRET acceptor signals.** (a) and (b) color coded trajectories lasting more than 200 ms found in one of the highlighted regions of Fig. 2b and c respectively. (c) Cumulative distribution of D values obtained on a single cell for EGFR dimers alone (red) and for the entire population of EGFR imaged in the donor channel (green).

Figure 1

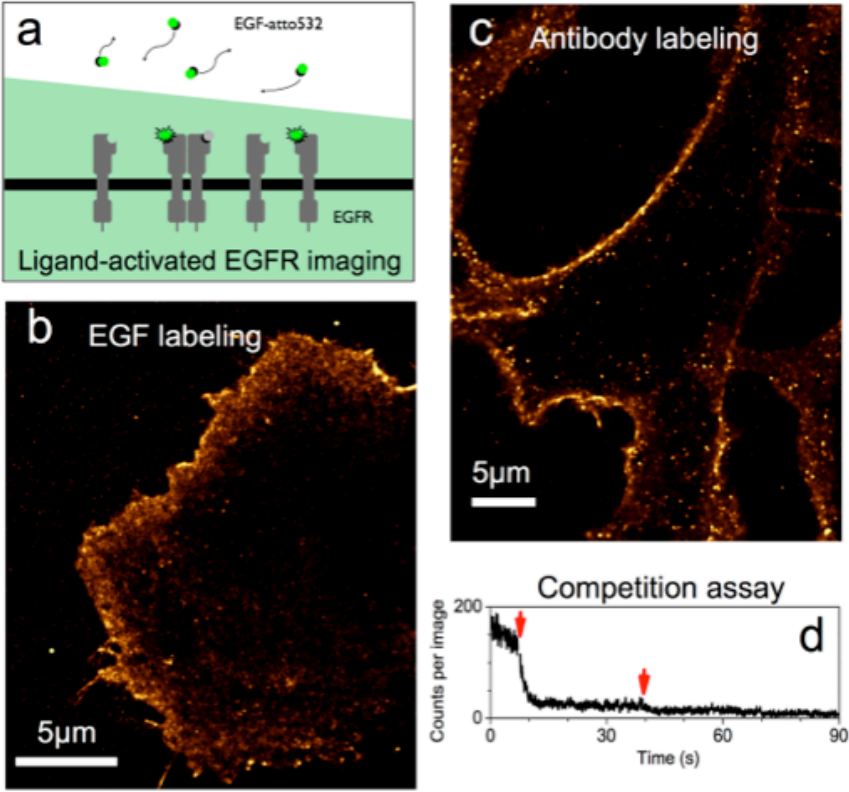

Figure 2

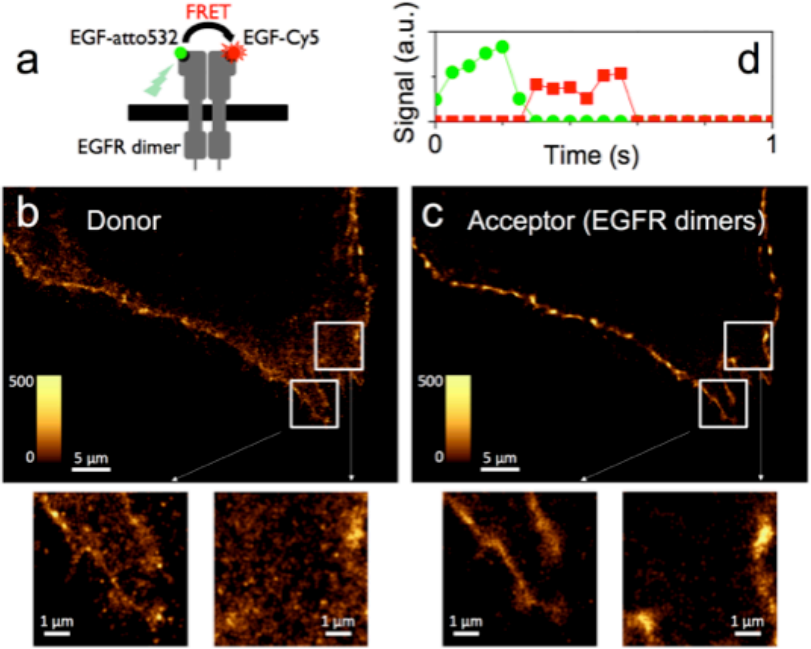

Figure 3

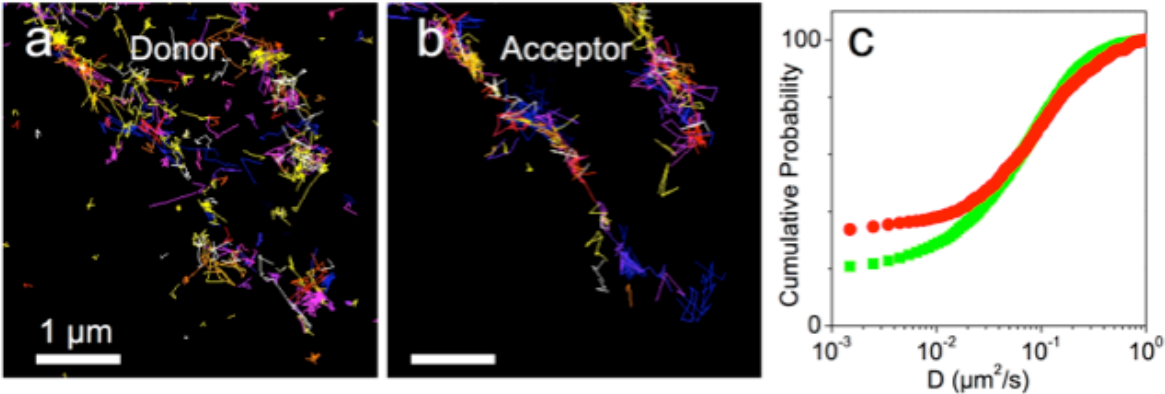